\documentclass[12pt]{iopart}
\usepackage{color}
\usepackage{cite}
\usepackage{braket}
\usepackage[pdftex]{graphicx}

\usepackage{iopams}  

\begin{document}

\title[]{Demonstration of the nearly continuous operation of an $^{171}$Yb optical lattice clock for half a year}

\author{Takumi Kobayashi$^{1}$, Daisuke Akamatsu$^{1}$, Kazumoto Hosaka$^{1}$, Yusuke Hisai$^{2}$, Masato Wada$^{1}$, Hajime Inaba$^{1}$, Tomonari Suzuyama$^{1}$, Feng-Lei Hong$^{2}$, Masami Yasuda$^{1}$}

\address{$^1$National Metrology Institute of Japan (NMIJ), National Institute of Advanced Industrial Science and Technology (AIST), 1-1-1 Umezono, Tsukuba, Ibaraki 305-8563, Japan\\
$^2$Department of Physics, Graduate School of Engineering Science, Yokohama National University, 79-5 Tokiwadai, Hodogaya-ku, Yokohama 240-8501, Japan\\}
\ead{takumi-kobayashi@aist.go.jp}
\vspace{10pt}
\begin{indented}
\item[]
\end{indented}

\begin{abstract}
Optical lattice clocks surpass primary Cs microwave clocks in frequency stability and accuracy, and are promising candidates for a redefinition of the second in the International System of Units (SI). However, the robustness of optical lattice clocks has not yet reached a level comparable to that of Cs fountain clocks which contribute to International Atomic Time (TAI) by the nearly continuous operation. In this paper, we report the long-term operation of an $^{171}$Yb optical lattice clock with a coverage of $80.3\%$ for half a year including uptimes of $93.9\%$ for the first 24 days and $92.6\%$ for the last 35 days. This enables a nearly dead-time-free frequency comparison of the optical lattice clock with TAI over months, which provides a link to the SI second with an uncertainty of low $10^{-16}$. By using this link, the absolute frequency of the $^{1}$S$_{0}-^{3}$P$_{0}$ clock transition of $^{171}$Yb is measured as $518\,295\,836\,590\,863.54(26)$ Hz with a fractional uncertainty of $5.0\times10^{-16}$. This value is in agreement with the recommended frequency of $^{171}$Yb as a secondary representation of the second.
\end{abstract}

%
%
%
%
%

\section{Introduction}
Several optical lattice clocks have achieved fractional frequency stabilities and uncertainties at the $10^{-18}$ level \cite{Bloom2014,Ushijima2015,Nicholson2015,McGrew2018,Bothwell2019}, which is better than primary Cs fountain microwave clocks. These achievements have stimulated discussion regarding a redefinition of the second in the International System of Units (SI) \cite{Gill2011,Hong2016,Riehle2018,Lodewyck2019}. 

With the aim of redefining the SI second, the absolute frequencies of optical lattice clocks have been measured by many groups and used to determine the recommended frequencies of secondary representations of the second \cite{Riehle2018}. The absolute frequency can be directly measured referenced to a Cs fountain clock if it is locally available in the laboratory \cite{Lodewyck2016,Lemke2009,Campbell2008,Targat2013,Grebing2016,Pizzocaro2017}. When a local Cs clock is not available, the absolute frequency can be measured using a satellite link to International Atomic Time (TAI). TAI is a global timescale computed by the Bureau International des Poids et Mesures (BIPM). BIPM provides the frequency difference between TAI and the SI second averaged over a month. Therefore, the continuous comparison of an optical clock with TAI for a month is desirable in terms of linking to the SI second without additional uncertainty resulting from the dead time. A long-term comparison is also needed to reduce the uncertainty arising from the satellite link \cite{Panfilo2010}. Since optical lattice clocks have commonly been operated intermittently, previous works have mostly employed stable local flywheels (e.g, an ensemble of hydrogen masers) that reach the $10^{-16}$ level to bridge the gaps in the operation of the lattice clocks \cite{Tanabe2015,Hachisu2017,McGrew2019,Pizzocaro2019}. A few groups have reported the nearly continuous operation of Sr optical lattice clocks with uptimes of $93\%$ for 10 days, $83\%$ for three weeks \cite{Lodewyck2016}, and $84\%$ for 25 days \cite{Hill2016}. 

The continuous operation of an optical lattice clock is essential for the calibration of the frequency of TAI as a secondary representation of the second \cite{Riehle2018} as well as the absolute frequency measurement. A robust lattice clock is also important for new applications including tests of fundamental physics \cite{Targat2013,McGrew2019,Blatt2008,Safronova2018,Derevianko2014,Wcislo2016,Takamoto2020}, relativistic geodesy \cite{Takano2016,Grotti2018}, and the generation of a stable local timescale \cite{Grebing2016,Hachisu2018,Milner2019,Yao2019}. 

The main challenge as regards realizing the continuous operation of an optical lattice clock is to stabilize the frequencies of several light sources including an optical frequency comb to allow the comparison of the optical clock and a microwave standard. So far, we have developed multi-branch erbium-doped-fiber-based frequency combs that can run continuously for long periods \cite{Inaba2006}. By stabilizing the frequencies of many light sources to the comb, we have constructed a simple and reliable laser system for clock experiments \cite{Hisai2019}. We have also developed a relocking scheme \cite{Kobayashi2019} for optical phase locking. These techniques should greatly assist the realization of the continuous operation. 

In this paper, we report the operation of an $^{171}$Yb optical lattice clock with a coverage of $80.3\%$ for half a year (185 days) including uptimes of $93.9\%$ for the first 24 days, $86.4\%$ for the second 27 days, $80.4\%$ for the third 30 days, $72.7\%$ for the fourth 35 days, $82.6\%$ for the fifth 25 days, and $92.6\%$ for the sixth 35 days. Using a single hydrogen maser with its stability limited by a flicker floor of $2\times10^{-15}$, we reduce the uncertainty in the link between the Yb clock and the SI second to the low $10^{-16}$ level, and measure the absolute frequency of the Yb clock transition. 

\section{Experimental setup}
\begin{figure}
\includegraphics[scale=0.44]{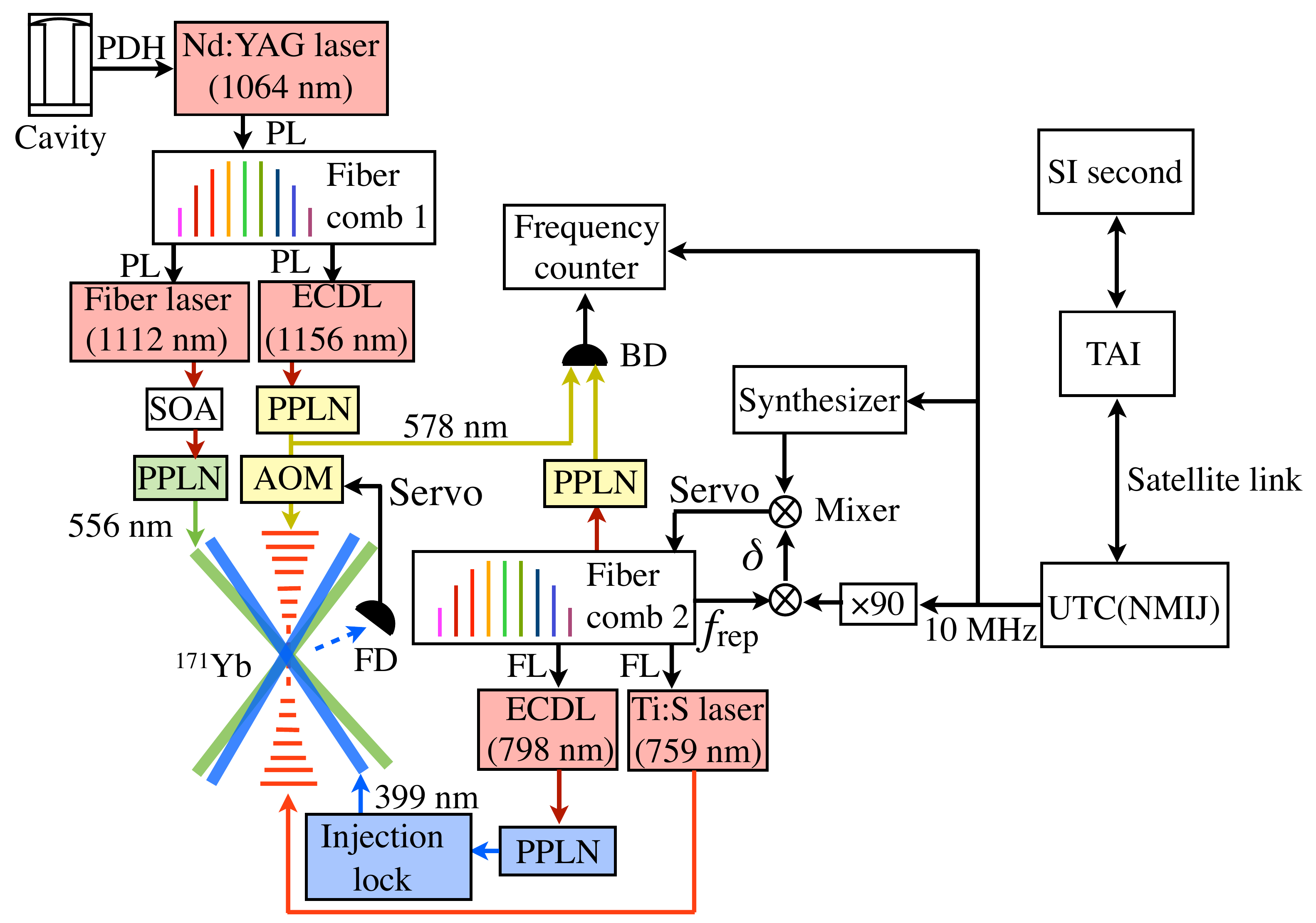}
\caption{Experimental setup. PDH: Pound-Drever-Hall stabilization, PL: Phase lock, ECDL: External cavity diode laser, PPLN: periodically poled LiNbO$_{3}$, Ti:S: Titanium sapphire, SOA: Semiconductor optical amplifier, FL: Frequency lock, BD: Beat detection, FD: Fluorescence detection.}
\label{experimentalsetup}
\end{figure}

Figure \ref{experimentalsetup} is a schematic diagram of the experimental setup. The setup for the Yb optical lattice clock is similar to that reported in our previous paper \cite{Kobayashi2018}. The clock utilizes the $^{1}$S$_{0}-^{3}$P$_{0}$ clock transition at 578 nm of $^{171}$Yb atoms trapped in an optical lattice operated at the magic wavelength \cite{Katori2003} of 759 nm. The 578 nm light was generated by the second harmonic generation (SHG) of an external cavity diode laser (ECDL) emitting at 1156 nm \cite{Kobayashi2016}. The lattice light at 759 nm was provided by a titanium sapphire (Ti:S) laser (M squared, SolsTiS-4000-SRX-R). Before loading the atoms to the optical lattice, two-stage laser cooling with the $^{1}$S$_{0}-^{1}$P$_{1}$ transition at 399 nm and the $^{1}$S$_{0}-^{3}$P$_{1}$ transition at 556 nm was carried out to decrease the temperature of the atoms to several tens of microkelvins. The 399 nm light was generated by an injection-locked 399 nm diode laser with an automatic relocking mechanism \cite{Saxberg2016}. A seed light for the injection locking was provided by the SHG of a 798 nm ECDL \cite{Kobayashi2016OE}. The 556 nm light for the second stage cooling was prepared by the SHG of a 1112 nm light from a fiber laser (NKT Photonics, Koheras) with a semiconductor optical amplifier. All the SHG processes were performed with single-pass periodically poled LiNbO$_{3}$ waveguides.

The 1112 nm fiber laser and the 1156 nm ECDL were phase locked to a narrow linewidth fiber comb (Fiber comb 1 in Fig. \ref{experimentalsetup}) \cite{Inaba2013,Iwakuni2012}. This comb was phase locked to a master Nd:YAG laser at 1064 nm, which was stabilized to an ultra-low expansion (ULE) cavity by using the Pound-Drever-Hall method. To compensate for the narrow capture range of optical phase locking, a recently-developed relocking scheme \cite{Kobayashi2019} was incorporated in the phase locking of the comb to the master laser and the phase locking of the 1112 nm and 1156 nm lasers to the comb. The 798 nm ECDL and the Ti:S laser were stabilized to another fiber comb (Fiber comb 2 in Fig. \ref{experimentalsetup}) by frequency locking with an electrical delay line \cite{Hisai2018}, which ensured a large capture range. Fiber comb 2 was referenced to the Coordinated Universal Time of the National Metrology Institute of Japan (UTC(NMIJ)), which is generated by a hydrogen maser and an auxiliary output generator to add an arbitrary frequency offset. 

The Yb optical lattice clock was operated with a cycle time of about 1.7 s. Yb atoms effused from an oven were decelerated with a Zeeman slower and cooled in first-stage cooling for about 1.2 s and then further cooled in second-stage cooling for 275 ms. In the present work, a window heated at 204$^{\circ}$C was installed inside a vacuum chamber in the pass of the Zeeman slower beam to prevent the atomic beam from coating the viewport. The atoms were then loaded into a vertically oriented one-dimensional optical lattice and spin-polarized with the $^{1}$S$_{0}-^{3}$P$_{1}$ transition at 556 nm for 20 ms. A Rabi $\pi$ pulse resonant on the clock transition was then applied for 40 ms. The atomic population and the excitation probability were deduced using a laser-induced fluorescence signal by the $^{1}$S$_{0}-^{1}$P$_{1}$ transition at 399 nm, and repumping on the $^{3}$P$_{0}-^{3}$D$_{1}$ transition at 1389 nm induced by a free-running distributed feedback laser. The excitation probability was used to calculate frequency corrections applied for an acousto-optic modulator (AOM), which steers the frequency of the clock laser to the atomic transition. The atomic transition was split into two Zeeman components $m_{F}=\pm1/2$ by applying a bias magnetic field of 65 $\mu$T. The clock laser was stabilized to each component.

Fiber comb 2 was also used to measure the frequency of the Yb clock against UTC(NMIJ). To transfer the stability of UTC(NMIJ) to the comb without degradation, we mixed the 7th harmonics of the repetition rate frequency $f_{\mathrm{rep}}\sim121.8$ MHz with a 900 MHz signal which was generated by the 90-fold frequency multiplication of a 10 MHz signal from UTC(NMIJ). The resulting output frequency from the mixer $\delta\sim47$ MHz was phase locked to a synthesizer referenced to UTC(NMIJ). After stabilizing the comb to UTC(NMIJ), the beat frequency between the clock laser and the comb was measured by using a zero dead-time frequency counter. 

\section{Demonstration of the nearly continuous operation}
\begin{figure}[h]
\includegraphics[scale=0.45]{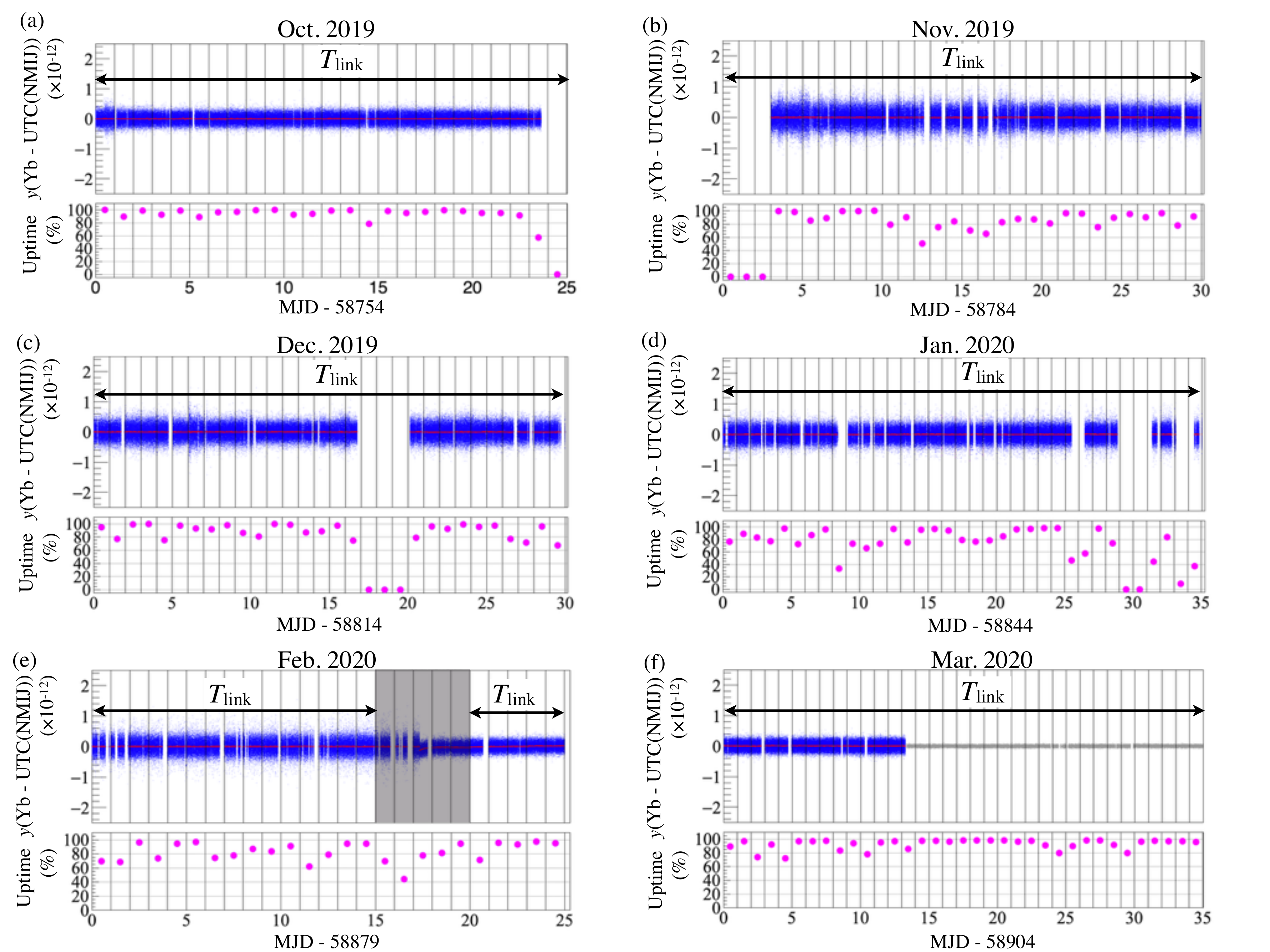}
\caption{Fractional frequency of the Yb clock relative to UTC(NMIJ) $(y(\mathrm{Yb}-\mathrm{UTC(NMIJ)}))$ and the uptime per day in the (a) first, (b) second, (c) third, (d) fourth, (e) fifth, and (f) sixth campaigns. The blue point corresponds to a 6.8 s average, the red point a $10^3$ s average, and the gray point in (f) a 30 s average. $T_{\mathrm{link}}$ indicates the period employed to calculate the frequency difference between UTC(NMIJ) and TAI. The data in the shaded region in (e) were not used for the calculation due to a large excursion of the phase of UTC(NMIJ) at MJD 58896.}
\label{continuousresult}
\end{figure}

The frequency of the Yb optical lattice clock was measured against UTC(NMIJ) during a half-year period from the Modified Julian Date (MJD) 58754 (28 September 2019) to MJD 58939 (31 March 2020). The Yb clock was operated with uptimes of 93.9$\%$ in the first 24-day campaign from MJD 58754 to MJD 58778, 86.4$\%$ in the second 27-day campaign from MJD 58787 to MJD 58814, 80.4$\%$ in the third 30-day campaign from MJD 58814 to MJD 58844, 72.7$\%$ in the fourth 35-day campaign from MJD 58844 to MJD 58879, 82.6$\%$ in the fifth 25-day campaign from MJD 58879 to MJD 58904, and 92.6$\%$ in the sixth 35-day campaign from MJD 58904 to MJD 58939 (see also Table 1). Figures \ref{continuousresult} (a) - (f) respectively show the fractional frequency differences between the Yb clock and UTC(NMIJ), denoted by $y(\mathrm{Yb}-\mathrm{UTC(NMIJ)})$, in the first - sixth campaigns. 
The uptime was calculated by using $N_{\mathrm{valid}}T_{\mathrm{cycle}}/T_{\mathrm{total}}$, where $N_{\mathrm{valid}}$ is the number of valid data points, $T_{\mathrm{cycle}}$ the clock cycle time ($\sim1.7$ s), and $T_{\mathrm{total}}$ the total period of the measurement. We discarded data that included the following events: (i) large excursions in the phase-locked frequencies that can occur during the relocking procedures; (ii) cycle slips in the frequency counting; and (iii) low excitation probabilities of the clock transition that lasted for a sufficiently long period. 

The fractional frequency difference $y(\mathrm{Yb}-\mathrm{UTC(NMIJ)})$ is given by
\begin{eqnarray}
y(\mathrm{Yb}-\mathrm{UTC(NMIJ)})&=&\frac{f^{\mathrm{a}}(\mathrm{Yb})}{f^{\mathrm{n}}(\mathrm{Yb)}}-\frac{f^{\mathrm{a}}(\mathrm{UTC(NMIJ))}}{f^{\mathrm{n}}(\mathrm{UTC(NMIJ)})}\nonumber\\
&\approx& \frac{f^{\mathrm{a}}(\mathrm{Yb})/f^{\mathrm{a}}(\mathrm{UTC(NMIJ))}}{f^{\mathrm{n}}(\mathrm{Yb)}/f^{\mathrm{n}}(\mathrm{UTC(NMIJ)})}-1,\quad
\label{yequalation}
\end{eqnarray} 
where $f^{\mathrm{a(n)}}(\mathrm{X})$ denotes the actual (nominal) frequency of X. The approximation is vaild when $(f^{\mathrm{a}}(X)-f^{\mathrm{n}}(X))\ll f^{\mathrm{n}}(X)$. Here we chose $f^{\mathrm{n}}(\mathrm{UTC(NMIJ)})=10$ MHz and  $f^{\mathrm{n}}(\mathrm{Yb)}=f^{\mathrm{CIPM}}(\mathrm{Yb})=518\,295\,836\,590\,863.6$ Hz, which is the CIPM (Comit\'e International des Poids et Mesures) recommended frequency of $^{171}$Yb \cite{Riehle2018}.

\begin{table}[h]
\caption{Uptimes of the Yb optical lattice clock}  
	\label{uptimetable}
	\begin{center} 
\begin{tabular}{cccc}
\hline
Campaign  & MJD & Period (days) & Uptime $(\%)$\\
\hline
1 & 58754 - 58778 & 24 & $93.9$ \\
2 & 58787 - 58814 & 27 & $86.4$ \\
3 & 58814 - 58844 & 30 & $80.4$ \\
4 & 58844 - 58879 & 35 & $72.7$ \\
5 & 58879 - 58904 & 25 & $82.6$ \\
6 & 58904 - 58939 & 35 & $92.6$ \\
\hline
Total & 58754 - 58939 & 185 & $80.3$ \\
\hline
\end{tabular}
\end{center}
\end{table}

The clock was mostly unattended during the full campaign period. Several experimental parameters were monitored remotely. A data acquisition computer automatically sent an email alert to operators when one of the experimental parameters was outside the normal range. In the first campaign (see Fig. \ref{continuousresult} (a)), to demonstrate almost dead-time-free operation ( $\ge90\%$ uptime per day), we always carried out manual tuning as soon as we received the email alert. Significant reductions in the uptime were caused by rare events such as the passage of a typhoon at MJD 58770 and the breaking of a diode laser used in fiber comb 1 at MJD 58778. In the second to sixth campaigns (see Figs. \ref{continuousresult} (b)-(f)), manual recovery was not always performed at night to demonstrate that a large uptime can also be achieved with minimum human effort. Large gaps from MJD 58831 to MJD 58834 in the third campaign were caused by the shutdown of the air conditioners for facility maintenance. In the fourth campaign, a diode laser used in fiber comb 2 was broken, which caused a large downtime from MJD 58873 to MJD 58875. At MJD 58877, we stopped the clock operation and carried out a systematic evaluation of the microwave synthesis (See Section \ref{uncertaintycombsect}).

\begin{figure}[h]
\includegraphics[scale=0.32]{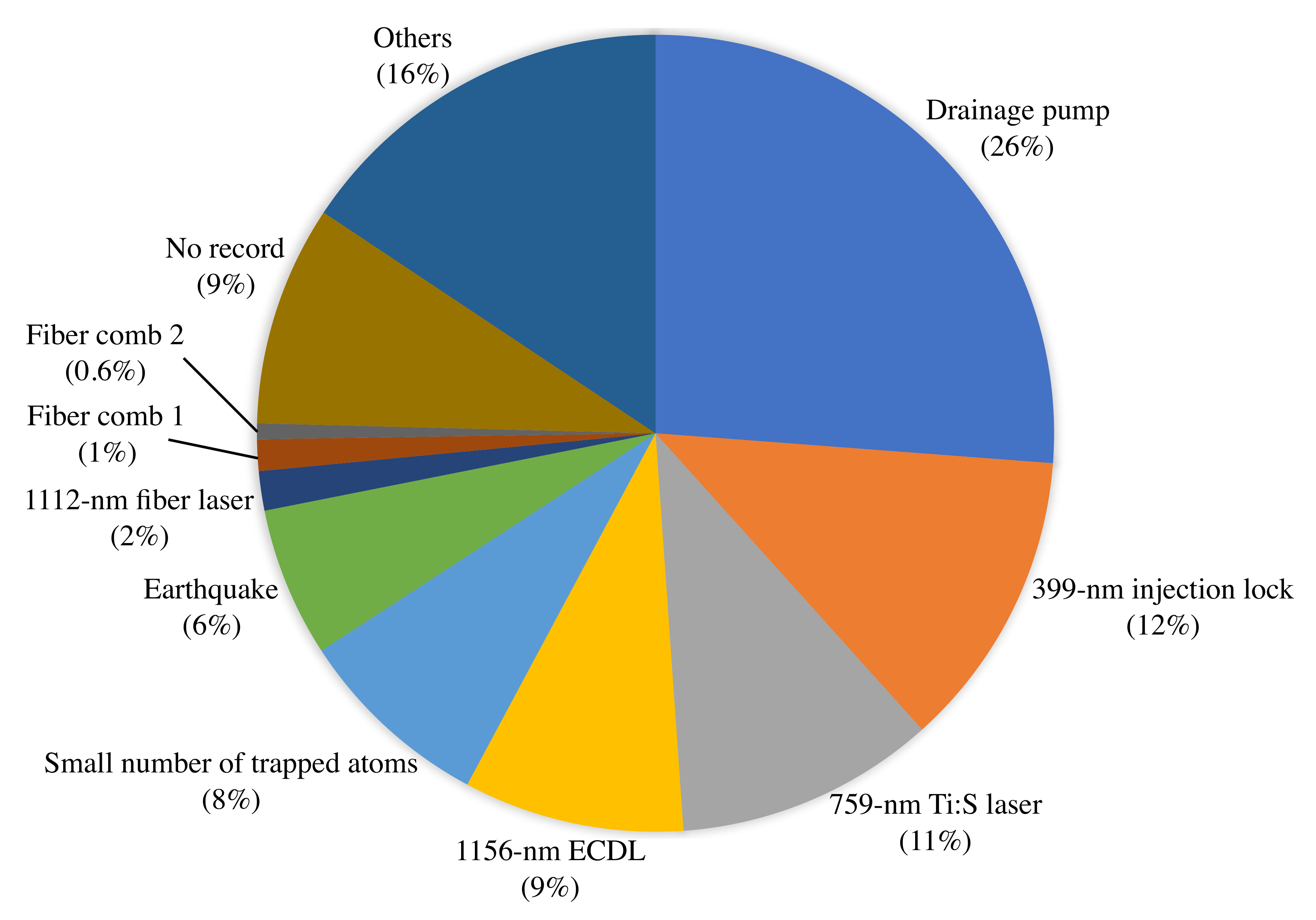}
\caption{Distribution of interrupting events during the full campaign period.}
\label{pichirt}
\end{figure}

Figure \ref{pichirt} shows the distribution of interrupting events that occurred during the full campaign period. The number of interrupting events were counted according to the email alert history. We could basically link the email history to the types of interruptions, but could not link some events due to a lack of records. These events are indicated by “No record” in Fig. \ref{pichirt}. The clock operation was mostly interrupted by acoustic noise from a drainage pump in the facility ($26\%$ of the total number of interruptions), the instability of the injection locking ($12\%$), mode hopping of the Ti:S laser ($11\%$) and the 1156 nm ECDL ($9\%$), and a small number of trapped atoms ($8\%$). Earthquakes ($6\%$) stopped the Yb clock mostly by dropping the lock of the 798 nm diode laser without implementing the relocking scheme. The other relockable light sources were resistant to small magnitude earthquakes. The other factors that interrupted the operation included rapid frequency drifts of the ULE cavity after the air conditioners were shutdown (MJD 58831 to MJD 58834) and temporary freezing of the data acquisition computer. 
\begin{figure}[h]
\includegraphics[scale=0.32]{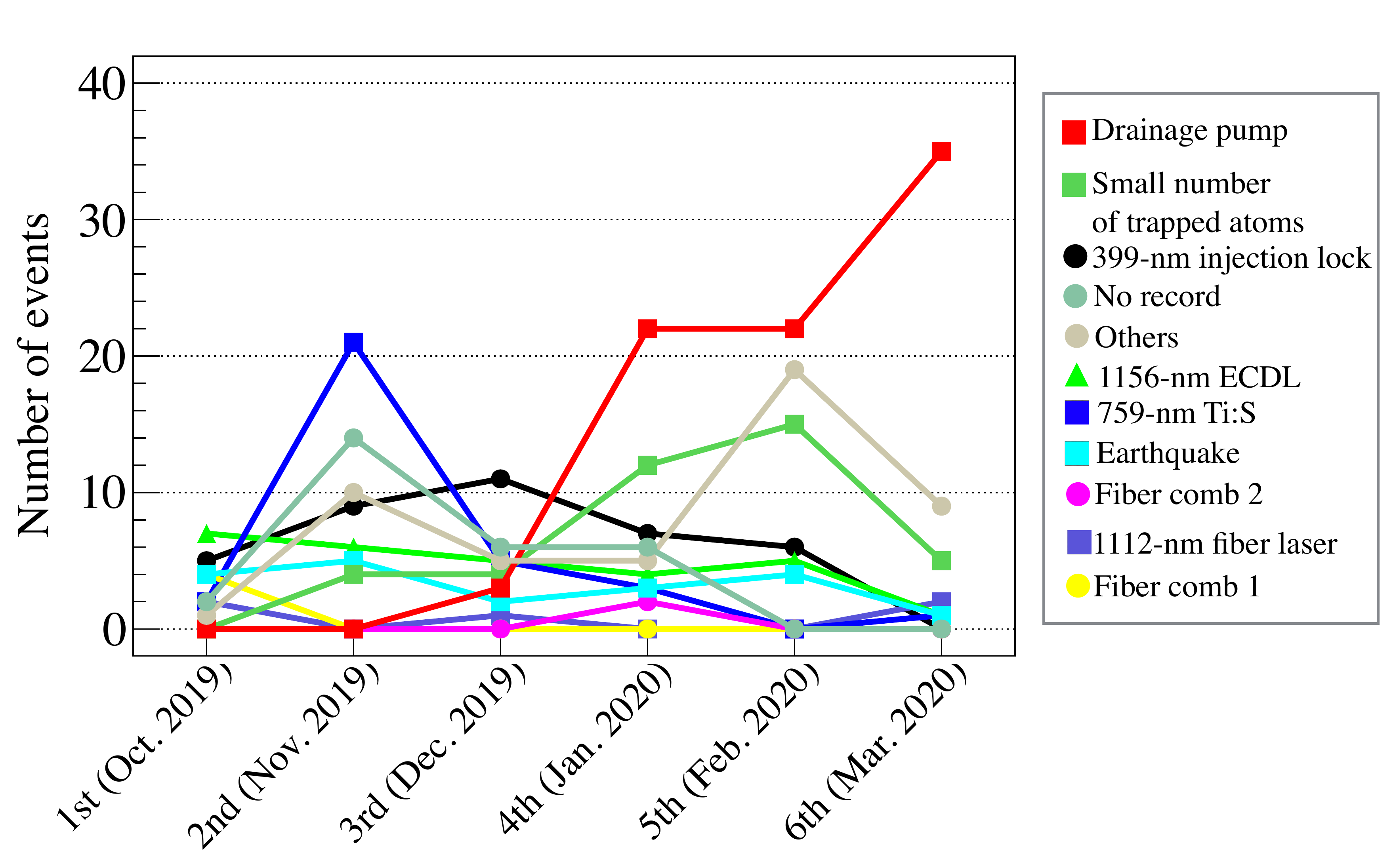}
\caption{The number of interrupting events in each campaign.}
\label{failhistory}
\end{figure}

Figure \ref{failhistory} shows the number of interrupting events in each campaign. Most of the interruptions occurred with a small number ($\lesssim10$) in each campaign except for the Ti:S laser (blue square), a small number of trapped atoms (green square), and a drainage pump (red square). The lock of the Ti:S laser failed frequently especially in the second campaign. This problem was partially solved in the following campaigns by cleaning the mirrors used to introduce the pump laser beam and improving the signal-to-noise ratio of the error signal for the frequency locking to the comb. From the second campaign, the number of atoms trapped in the optical lattice gradually decreased (see Section \ref{uncertaintyevaluationsect}), which increased the occurrence of lock failures. From the end of the third campaign, the Yb clock was frequently interrupted by a drainage pump, which had been installed in the facility during the third campaign. The pump generated acoustic noise for about 30 minutes once a day when it drained water. During this period, the excitation probability of the atomic transition became almost zero due to the broadening of the clock laser linewidth.

\section{Frequency stability and uncertainty of the optical lattice clock}
\label{uncertaintyevaluationsect}
\begin{figure}[h]
\includegraphics[scale=0.4]{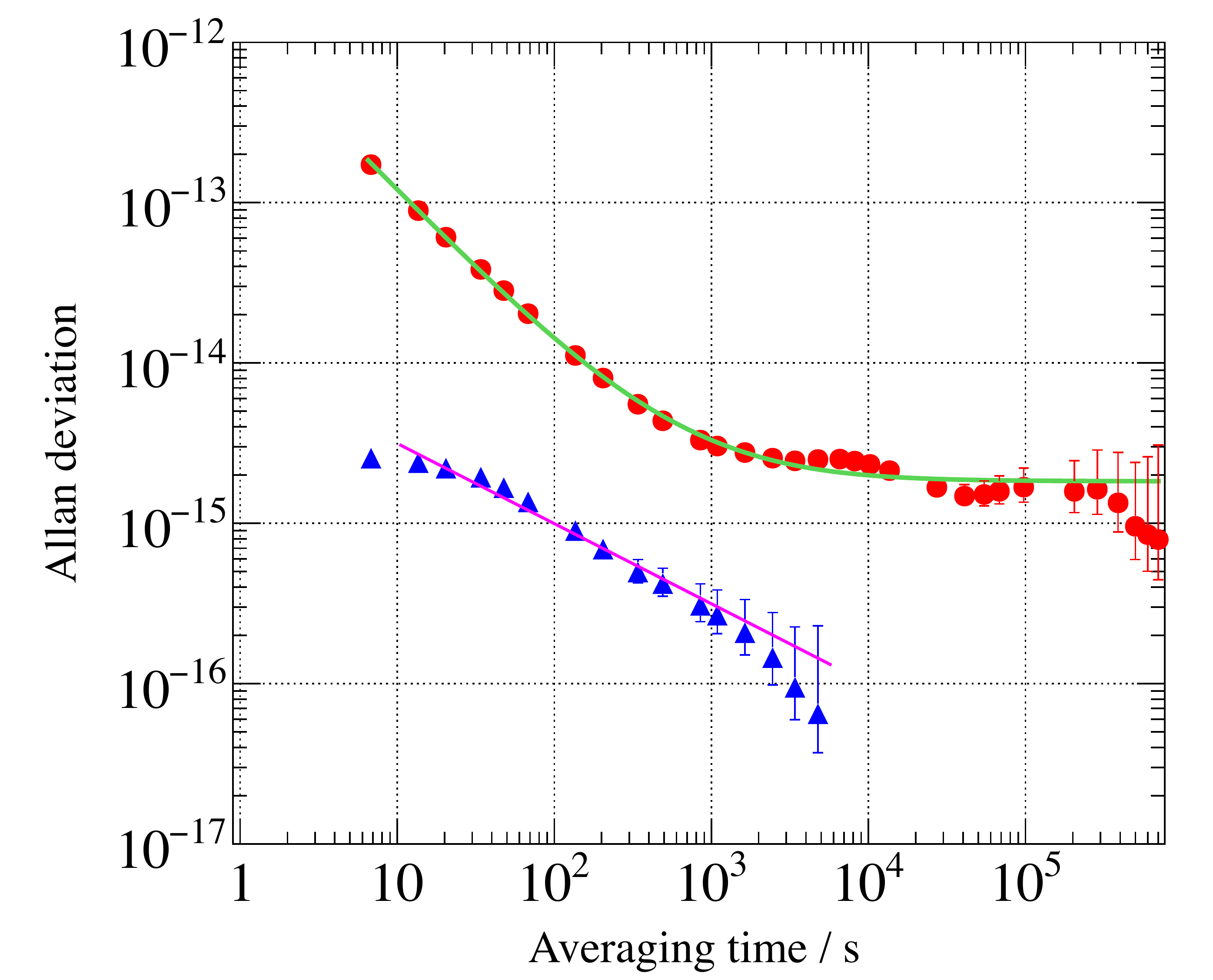}
\caption{Allan deviation of UTC(NMIJ) referenced to the Yb clock (red dots) and the frequency ratio Yb/Sr (blue triangles). The green curve shows the stability of a simulated model of UTC(NMIJ) used to calculate the dead time uncertainty (see text). The purple line indicates a slope of $1.0\times10^{-14}/\sqrt{(\tau/\mathrm{s})}$.}
\label{allandeviation}
\end{figure} 

\begin{table}[h]
\caption{Systematic shifts and uncertainties of the $^{171}$Yb optical lattice clock obtained in the first campaign from MJD 58754 to MJD 58778.}  
	\label{systematictalbe}
	\begin{center} 
\begin{tabular}{lcc}
\hline
Effect  & Shift & Uncertainty \\
&($\times10^{-17}$) & ($\times10^{-17}$) \\
\hline
Lattice light & 3.4 & 33.1 \\
BBR & $-263.8$ & 20.8 \\
Density & $-8.3$ & 6.4 \\
Second-order Zeeman & $-5.2$ & 0.3 \\
Probe light & 0.4 & 0.2 \\
Servo error & $-4.7$ & 1.1\\
AOM switching & $-$& 1 \\
\hline
Total & $-278.3$ & 39.6\\
\hline
\end{tabular}
\end{center}
\end{table}
\begin{figure}[h]
\includegraphics[scale=0.45]{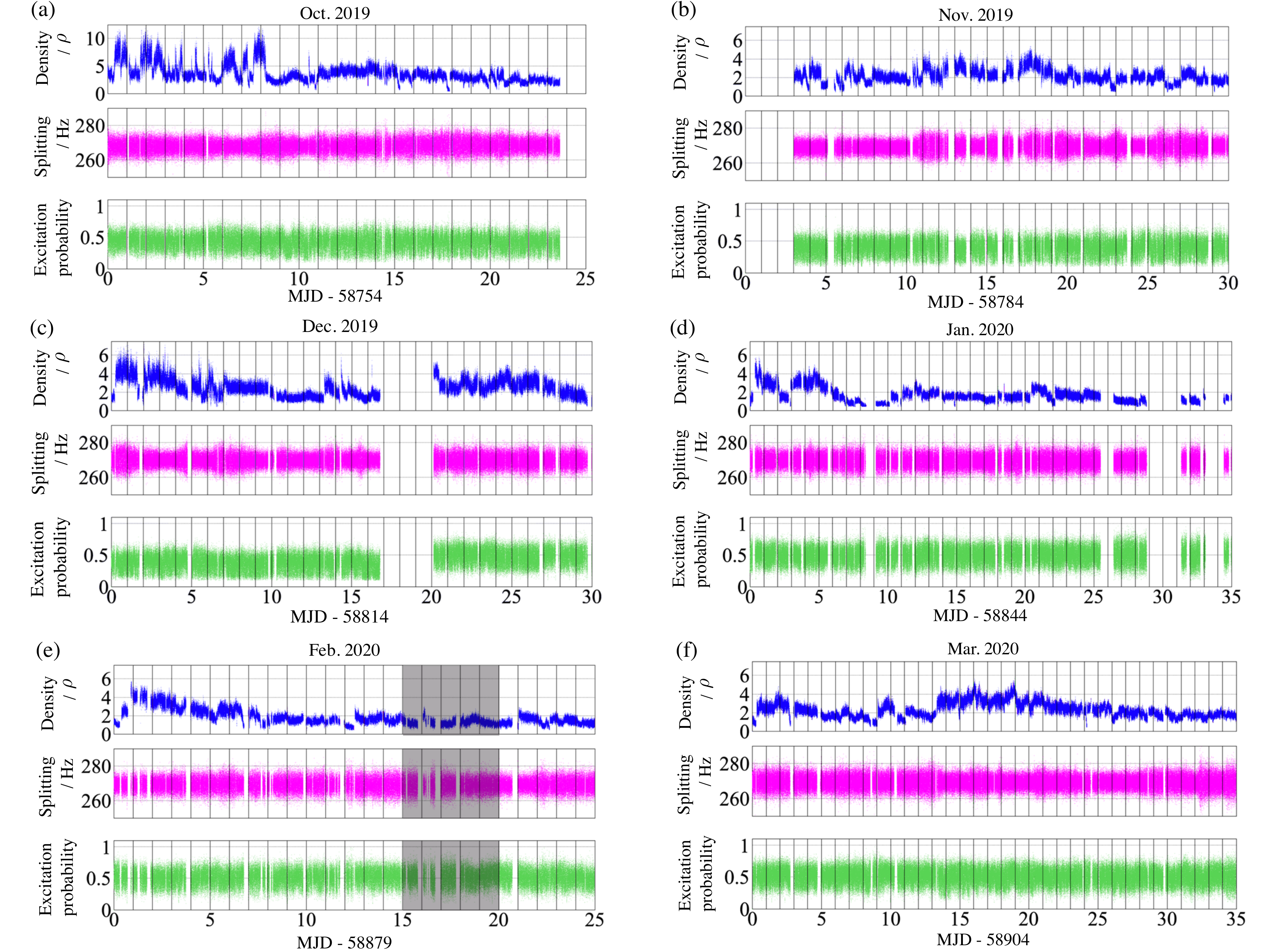}
\caption{Atomic density ($\rho\sim10^{14}$ m$^{-3}$) in the optical lattice (blue points), Zeeman splitting (pink points), excitation probability (green points) in the (a) first, (b) second, (c) third, (d) fourth, (e) fifth, and (f) sixth campaigns. The data point corresponds to a 6.8 s average. Some of the data at MJD 58789 in (b) were not recorded. The data in the shaded region (MJD 58894 to MJD 58899 in (e)) were not used for the systematic evaluation. The fiber for the 399 nm light was replaced at MJD 58794 in (b), MJD 58814 in (c), MJD 58879 in (e),  and MJD 58917 in (f).  The 1112 nm semiconductor amplifier was replaced at MJD 58844 in (d), MJD 58853 in (d), MJD 58891 in (e), and MJD 58913 in (f).}
\label{clockdata}
\end{figure}
We evaluated the frequency stability of the Yb optical lattice clock by comparing it with our Sr optical lattice clock \cite{Akamatsu2014}. Figure \ref{allandeviation} shows the Allan deviation of the frequency ratio Yb/Sr, which was improved with a slope of $1.0\times10^{-14}/\sqrt{(\tau/\mathrm{s})}$ and reached low $10^{-16}$ in several thousand seconds.  Figure \ref{allandeviation} also shows the frequency stability of UTC(NMIJ) relative to the Yb clock, which was calculated from the data of the first campaign (see Fig. \ref{continuousresult} (a)). During this campaign period, the frequency offset of the auxiliary output generator for UTC(NMIJ) was constant.

Table \ref{systematictalbe} lists the systematic shifts and uncertainties of the Yb optical lattice clock obtained in the first campaign. The uncertainty evaluation is mostly based on our previous evaluation \cite{Kobayashi2018}. The density shift and the servo error differed slightly depending on the data obtained in each campaign.

The lattice induced light shift was estimated using a model \cite{Katori2015,Nemitz2016,Nemitz2019} that includes the contributions from the dominant electric-dipole ($E1$) polarizability, the multipolar ($M1$ and $E2$) polarizability, and hyperpolarizability \cite{Brown2018}. To calculate the shift using this model, the effective trap potential depth $V_{\mathrm{e}}$, which is less than the maximum trap depth $V_{0}=450E_{r}$ due to the radial motion of the atoms, and the average vibrational quantum number $\Braket{n}$ were estimated by sideband spectroscopy of the clock transition \cite{Blatt2009}. The obtained values were $V_{\mathrm{e}}=342(28)E_{r}$ and $\Braket{n}=2.1(6)$. The frequency of the lattice laser was stabilized at $\nu_{\mathrm{l}}=394\,798\,263$ MHz. 

To find the $E1$ polarizability coefficient $a$ and the $E1$ magic frequency $\nu_{\mathrm{E1}}$, we have previously measured the light shift as a function of the frequency of the lattice laser \cite{Kobayashi2018}, and fitted the obtained data with the light shift model \cite{Nemitz2016}. To estimate the variation of $a$ and $\nu_{\mathrm{E}}$ resulting from the systematic uncertainty of the fixed parameters in the light shift model ($V_{e}$, $\Braket{n}$, and the multipolar and hyperpolarizability shifts), we have employed a Monte Carlo method in which the fittings were repeated with different permutations of the fixed parameters. Since the distribution of $a$ obtained from many fittings was asymmetric and had a very long tail mainly due to a relatively large uncertainty in our estimation of $V_{\mathrm{e}}$, we have previously calculated a mean value of $a$ by removing rare events in the tail. Here, we employed a value of $a$ that was determined with center values of the fixed parameters. The uncertainty of $a$ was estimated from a $68\%$ coverage interval of the distribution of $a$. The same method was applied to the estimation of $\nu_{\mathrm{E1}}$. We obtained $a=0.020(13)$ mHz/(MHz$E_{r}$) and $\nu_{E1}=394\,798\,244(25)$ MHz, which agreed with our previous estimation \cite{Kobayashi2018} and the values reported by other groups \cite{Nemitz2019,Brown2018,Kim2017,Pizzocaro2019}. Finally, the light shift was estimated to be $3.4(33.1)\times10^{-17}$ with $a$, $\nu_{E1}$, $\nu_{\mathrm{l}}$, $\Braket{n}$, $V_{\mathrm{e}}$ estimated here, and the sensitively coefficients for the multipolar and hyperpolarizability shifts used in our previous evaluation \cite{Kobayashi2018}.

The blackbody radiation (BBR) shift was estimated to be $-263.8(20.8)\times10^{-17}$, which includes the contribution from the vacuum chamber and the atomic oven considered in our previous evaluation \cite{Kobayashi2018} and also that from the newly-installed heated window at 204$^{\circ}$C. To include this new contribution, the BBR shift induced by the heated window was calculated using a model \cite{Middelmann2011} in which the atoms are located at the center of a stainless-steel sphere with an emissivity of 0.1. The radius of the sphere is equal to the distance between the atoms and the heated window ($\sim140$ mm). The BBR photons at 204$^{\circ}$C are provided from a small portion of the sphere's surface with its area equal to that of the heated window seen by the atoms ($\sim260$ mm$^{2}$). This model takes into account the diffuse reflection of the BBR photons on the stainless-steel wall, which increases the effective solid angle by a factor of $\sim10$ compared with the geometric solid angle. With the sensitivity coefficient \cite{Sherman2012,Beloy2014}, the calculated shift due to the heated window was $-1.8\times10^{-16}$. This shift value was also conservatively taken as an uncertainty.

The density shift was estimated from the atomic density in each campaign. The blue points in Fig. \ref{clockdata} show the atomic density data. The density decreased with a time scale of $\sim10$ days largely due to a reduction in the transmission efficiency of the 399 nm light through the optical fiber and the output power of the 1112 nm semiconductor amplifier. Some of these components were replaced (see the caption of Fig. \ref{clockdata}) when we could not keep the stabilization of the clock laser to the atomic transition. The status of the stabilization was confirmed by monitoring the Zeeman splittings and the excitation probabilities (see pink and green points in Fig. \ref{clockdata}). The short-term (a few days) fluctuation of the density was mainly caused by variations in the spectral purity of the injection locking, the polarization rotation of the cooling lasers through the optical fiber, and pointing instabilities of the cooling beam at 399 nm induced by an AOM or a damaged optical fiber. In the first campaign (see blue points in Fig. \ref{clockdata} (a)), the mean value of the atomic density was $3.6(1.6)\rho$ with $\rho\sim10^{14}$ m$^{-3}$. The uncertainty of $1.6\rho$ was estimated from the standard deviation. We also added an uncertainty of $\sim10\%$ to the estimation of the density according to the fluctuation of the 399 nm probe power. With our sensitivity coefficient \cite{Kobayashi2018} and a scaling of the trap volume with $V_0^{-3/2}$ \cite{Nicholson2015}, the shifts were determined as $-8.3(6.4)\times10^{-17}$, $-4.9(3.4)\times10^{-17}$, $-5.7(4.1)\times10^{-17}$, $-3.9(3.0)\times10^{-17}$, $-4.5(3.5)\times10^{-17}$, and $-5.3(3.7)\times10^{-17}$ in the first, second, third, fourth, fifth, and sixth campaigns, respectively. We observed that the mean excitation probability changed from 0.42 to 0.50 after the air conditioners were shutdown (MJD 58831 to MJD 58834) (see Fig. \ref{clockdata} (c)). We attributed this to a variation of the spin-polarization efficiency due to a large frequency change of the ULE cavity. This can change the shift coefficient, since the density shift depends on (i) the excitation probability mostly due to the $p-$wave interaction \cite{Lemke2011,Ludlow2011,Lee2016} and (ii) the impurity of the spin-polarization \cite{Ludlow2011,McGrew2018}. Based on the measured results of Ref. \cite{Ludlow2011}, we estimated that in our relatively low atomic density, the change due to those effects is negligibly small compared with the uncertainty in our evaluation.

The second-order Zeeman shift was estimated from the Zeeman splitting measured in each campaign (see pink points in Fig. \ref{clockdata}). In the first campaign, the mean value of the splitting was 268(3) Hz, where the uncertainty was estimated from the standard deviation. The shift was found to be $-5.2(3)\times10^{-17}$ using our sensitivity coefficient \cite{Kobayashi2018}. The mean splitting value was almost constant (268 - 270 Hz) during the entire campaign period. 

The probe light shift was calculated to be $4(2)\times10^{-18}$ from the shift value of Ref. \cite{McGrew2018} and the fact that the laser intensity adjusted for the $\pi$ pulse was inversely proportional to the time duration of the pulse.

The servo error was estimated by averaging the differences between the excitation probabilities of the high- and low-spectral shoulders for the data obtained in each measurement campaign. The servo errors were estimated to be $-4.7(1.1)\times10^{-17}$, $-4.8(1.0)\times10^{-17}$, $4.6(10.8)\times10^{-17}$, $-2.8(1.2)\times10^{-17}$, $-6.0(0.9)\times10^{-17}$, and $-4.1(1.4)\times10^{-17}$ in the first, second, third, fourth, fifth, and sixth campaigns, respectively. The relatively large uncertainty of $10.8\times10^{-17}$ in the third campaign arose from the rapid frequency drifts of the ULE cavity after the air conditioners were shutdown (MJD 58831 to MJD 58834).

\section{Uncertainty of the frequency comparison by the comb}
\label{uncertaintycombsect}
\begin{figure}[h]
\includegraphics[scale=0.4]{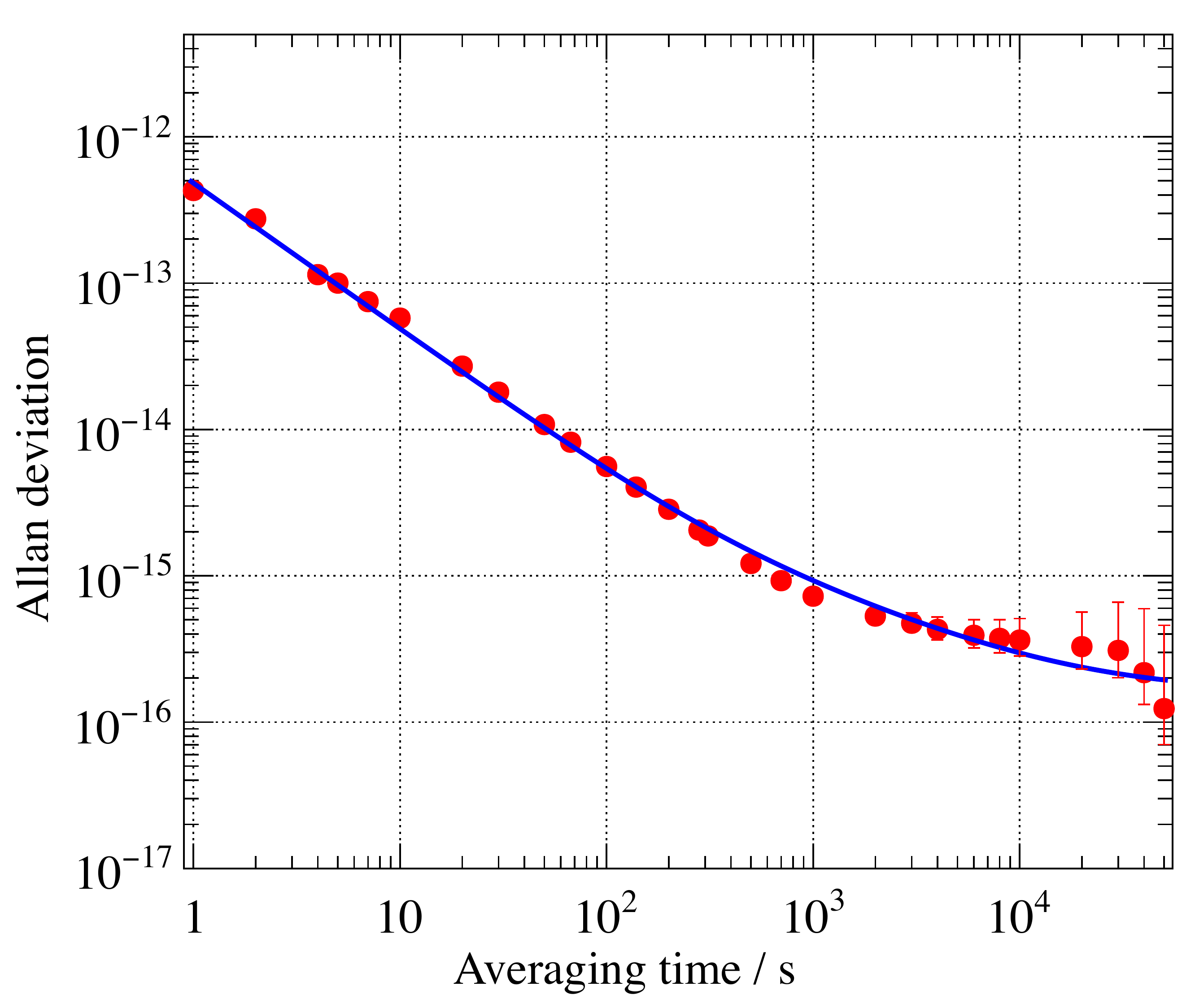}
\caption{Allan deviation calculated from the frequency difference between two independent combs except for the referece signal of UTC(NMIJ). The blue curve indicates the fit of a function which includes the white phase, the white frequency, and the flicker frequency noise components.}
\label{combcomb}
\end{figure}
To evaluate the uncertainty of the comb-based frequency comparison between the Yb clock and UTC(NMIJ), we compared fiber comb 2 (see Fig. \ref{experimentalsetup}) with another comb (fiber comb 3, not shown in Fig. \ref{experimentalsetup}) that was independent except for the 10 MHz reference signal of UTC(NMIJ). These combs had independent setups for the microwave synthesis including the 90-fold frequency multiplication. We simultaneously measured the frequency of an acetylene-stabilized laser at 1.5 $\mu$m against UTC(NMIJ) by using these two combs, and calculated the frequency difference between the combs. In the first measurement from MJD 58712 to MJD 58713, the fractional frequency difference (fiber comb 2 - fiber comb 3) was found to be $9.7\times10^{-17}$, which was the average of the frequency difference data obtained for $\sim10^{5}$ s. Figure \ref{combcomb} shows the Allan deviation calculated from the frequency difference data. The Allan deviation is likely limited by a flicker floor of low $10^{-16}$ at averaging times $\tau\ge10^{4}$ s. By fitting the Allan deviation with a function which includes the white phase, the white frequency, and the flicker frequency noise components, the flicker component was estimated to be $1.6\times10^{-16}$. In the second measurement at MJD 58877, the frequency difference and the flicker component were $-1.1\times10^{-16}$ and $4.5\times10^{-16}$, respectively. We took an average of the two flicker noise values ($3.1\times10^{-16}$) and divided it by $\sqrt{2}$. The obtained value $2.2\times10^{-16}$ was adopted as an uncertainty of the frequency comparison. Previous optical-optical comparisons \cite{Nakajima2010} have shown that the uncertainty resulting from the comb itself is negligibly small. We also carried out another measurement in which the two combs shared the common frequency multiplier, and observed that the Allan deviation was improved with a $\tau^{-1}$ slope for $\tau>10^{3}$ s. Thus, we attributed the flicker noise mainly to the frequency multiplier.

\section{Absolute frequency measurement}
\begin{figure}[h]
\includegraphics[scale=0.42]{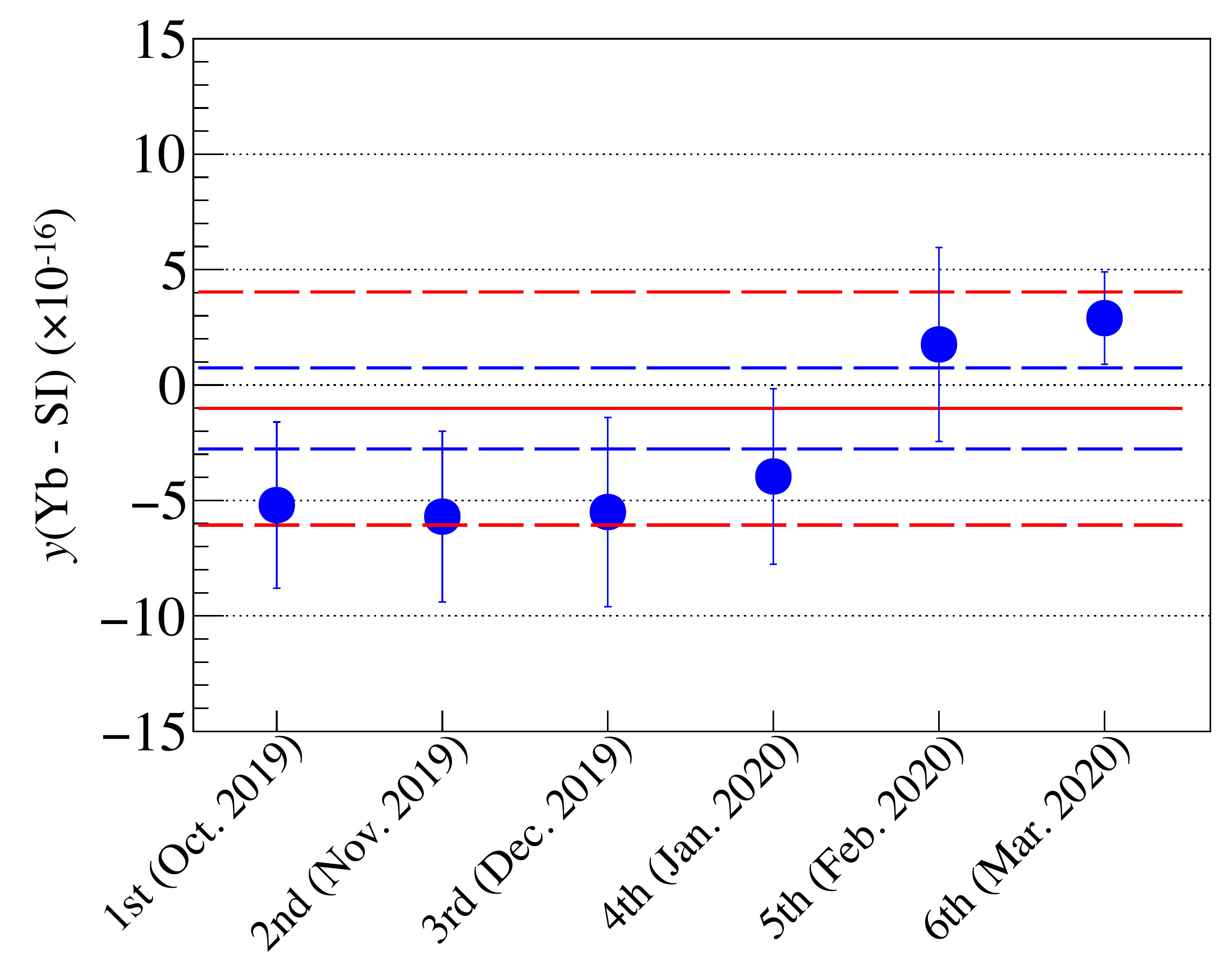}
\caption{Fractional frequency of the Yb clock referenced to the SI second $(y(\mathrm{Yb}-\mathrm{SI}))$ obtained in each campaign. The error bar is given by the total statistical uncertainty for each campaign. The solid red line indicates the weighted mean, the dashed blue lines indicate the total statistical uncertainty in the full period, and the dashed red lines indicate the total uncertainty including the systematic uncertainties.}
\label{reprocibility}
\end{figure}

Since UTC(NMIJ) was continuously compared with TAI via a satellite link (see Fig. \ref{experimentalsetup}), the frequency of the Yb clock referenced to the SI second $y(\mathrm{Yb}-\mathrm{SI})$ can be deduced from the data $y(\mathrm{Yb}-\mathrm{UTC(NMIJ)})$ in Fig. \ref{continuousresult} by the relationship,
\begin{eqnarray}
y(\mathrm{Yb}-\mathrm{SI})&=&y(\mathrm{Yb}-\mathrm{UTC(NMIJ)}) \nonumber\\
&&+y(\mathrm{UTC(NMIJ)}-\mathrm{TAI})\nonumber\\
&&+y(\mathrm{TAI}-\mathrm{SI}), 
\label{equation}
\end{eqnarray}
where $y(\mathrm{UTC(NMIJ)}-\mathrm{TAI})$ and $y(\mathrm{TAI}-\mathrm{SI})$ denote the fractional frequency differences between (i) UTC(NMIJ) and TAI and (ii) TAI and SI, respectively. These values are provided in Circular T \cite{circulart}, which is a monthly report issued by BIPM. Since BIPM only computes $y(\mathrm{UTC(NMIJ)}-\mathrm{TAI})$ for 5-day intervals, we chose a 25-day period (MJD 58754 - MJD 58779), a 30-day period (MJD 58784 - MJD 58814), a 30-day period (MJD 58814 - MJD 58844), a 35-day period (MJD 58844 - MJD 58879), a 20-day period (MJD 58879 - MJD 58884 and MJD 58899 - MJD 58904), and a 35-day period (MJD 58904 - MJD 58939) for the first - sixth campaigns, respectively, to calculate the average value of $y(\mathrm{UTC(NMIJ)}-\mathrm{TAI})$. These periods are indicated by $T_{\mathrm{link}}$ in Fig. \ref{continuousresult}. In the fifth campaign, a 5-day period from MJD 58894 to MJD 58899 (see the shaded region of Fig. \ref{continuousresult} (e)) was excluded, since a large excursion of the phase of UTC(NMIJ) occurred during maintenance of the hydrogen maser at MJD 58896. Figure \ref{reprocibility} shows the $y(\mathrm{Yb}-\mathrm{SI})$ value obtained for each campaign. 

\begin{table}[h]
\caption{Uncertainty budget for the absolute frequency measurement of the $^{171}$Yb clock transition.}  
	\label{clockbudget}
	\begin{center} 
\begin{tabular}{lcc}
\hline
Effect  & First campaign & Full period \\
&($\times10^{-16}$) & ($\times10^{-16}$) \\
\hline
Statistics & &\\
Yb & $0.07$ & $0.03$ \\
Dead time in Yb - UTC(NMIJ) & 2.0 & 0.6 \\
Dead time in TAI - SI & 1.6 & 0.4 \\
UTC(NMJI) - TAI satellite link & 2.3 & 0.8 \\
TAI - SI & 0.9 & 0.4 \\
\hline
Statistics total & 3.6 & 1.8\\
\hline
Systematics & &\\
Yb & $4.0$& $4.0$ \\
Microwave synthesis & 2.2 & 2.2\\
TAI - SI  & 1.1 & 1.1 \\
Gravitational red shift & 0.6 & 0.6 \\
\hline
Systematics total & 4.7 & 4.7 \\
\hline
Total & 5.9 & 5.0\\
\hline
\end{tabular}
\end{center}
\end{table}

The choice of the period $T_{\mathrm{link}}$ causes an uncertainty due to the dead time in the comparison of the Yb clock and UTC(NMIJ). This dead time uncertainty was calculated by using a numerical simulation described in Ref. \cite{Yu2007}. The simulation generated time series data of the frequency according to the noise characteristics: $1\times10^{-12}/(\tau/\mathrm{s})$ for the white phase modulation, $7\times10^{-14}/\sqrt{(\tau/\mathrm{s})}$ for the white frequency modulation (FM), $2\times10^{-15}$ for the flicker FM, and $4\times10^{-24}\sqrt{(\tau/\mathrm{s})}$ for the random walk FM. These noise parameters were chosen to reproduce the typical stability of UTC(NMIJ) (see green curve in Fig. \ref{allandeviation}). We generated two hundred time series patterns and calculated the standard deviation between the frequency averaged over the entire 25-, 30-, or 35-day period and the mean frequency for the time during which the Yb clock was operated. This standard deviation corresponds to the dead time uncertainty. The dead time uncertainties were estimated as $2.0\times10^{-16}$, $3.0\times10^{-16}$, $3.3\times10^{-16}$, $3.3\times10^{-16}$, $1.1\times10^{-16}$, and $8\times10^{-17}$ for the first, second, third, fourth, fifth, and sixth campaigns, respectively. A frequency correction arose from the dead time when the frequency steering of UTC(NMIJ) was carried out. We added corrections of $-9.9(2)\times10^{-18}$, $-1.0(9)\times10^{-16}$, and $7.9(5)\times10^{-17}$ in the third, fifth, and sixth campaigns, respectively.

In addition, the uncertainty resulting from the dead time when we compare TAI and SI arises from the fact that BIPM calculates the $y(\mathrm{TAI}-\mathrm{SI})$ value based on an evaluation with primary and secondary frequency standards for 25, 30, or 35 days, whereas our measurement period $T_{\mathrm{link}}$ did not fully cover the evaluation period of BIPM in the first and fifth campaigns. This uncertainty was estimated to be $1.6\times10^{-16}$ in the first campaign, and $4\times10^{-17}$ in the full period based on the stability of the Echelle Atomique Libre (EAL) given in Circular T: $1\times10^{-15}/\sqrt{(\tau/\mathrm{day})}$ for the white FM, $0.35\times10^{-15}$ for the flicker FM, and $0.2\times10^{-16}\sqrt{(\tau/\mathrm{day})}$ for the random walk FM.

The satellite link uncertainty in $y(\mathrm{UTC(NMIJ)}-\mathrm{TAI})$ was calculated to be $2.3\times10^{-15}$ for $T_\mathrm{link}=25$ days, $2.0\times10^{-16}$ for $T_{\mathrm{link}}=30$ days, and $1.7\times10^{-16}$ for $T_{\mathrm{link}}=35$ days using the recommended formula $(\sqrt{2}u_{\mathrm{A}})/[(3600\times24\times5)(\frac{T_\mathrm{link}}{5})^{0.9}]$ \cite{Panfilo2010} with the statistical uncertainty $u_{\mathrm{A}}=0.3$ ns reported in Circular T. 

The statistical and systematic uncertainties of $y(\mathrm{TAI}-\mathrm{SI})$ were calculated from the total uncertainty of $y(\mathrm{TAI}-\mathrm{SI})$ and the systematic uncertainties of the primary and secondary frequency standards reported in Circular T. During the present half-year period, the frequency of TAI was calibrated by using the following frequency standards: Cs thermal beam clocks (PTB-Cs1 and PTB-Cs2 \cite{Bauch2005}), Cs fountain clocks (SYRTE-FO1, SYRTE-FO2, SYRTE-FOM \cite{Guena2012}, PTB-CSF1, PTB-CSF2 \cite{Weyers2018}, SU-CsFO2 \cite{Blinov2017}, and NIM5 \cite{Fang2015}), a Rb fountain clock (SYRTE-FORb \cite{Guena2014}), and a Sr optical lattice clock (NICT-Sr1 \cite{Hachisu2018}). Since Circular T only provides the total uncertainty of $y(\mathrm{TAI}-\mathrm{SI})$, the systematic uncertainty of $y(\mathrm{TAI}-\mathrm{SI})$ was estimated using the systematic uncertainties and weights of the individual frequency standards based on a method described in Ref. \cite{Hachisu2016}. The estimated systematic uncertainty of $y(\mathrm{TAI}-\mathrm{SI})$ changed between $1.1\times10^{-16}$ and $1.2\times10^{-16}$ during the half year period. We adopted an average value of $1.1\times10^{-16}$ for the full campaign period. The statistical uncertainty of $y(\mathrm{TAI}-\mathrm{SI})$ was calculated with the above systematic uncertainty to reproduce the total uncertainty given in Circular T. The statistical uncertainty ranged between $0.8\times10^{-16}$ and $1.5\times10^{-16}$, and decreased to $4\times10^{-17}$ in the full campaign period.

\begin{figure}[t]
\includegraphics[scale=0.45]{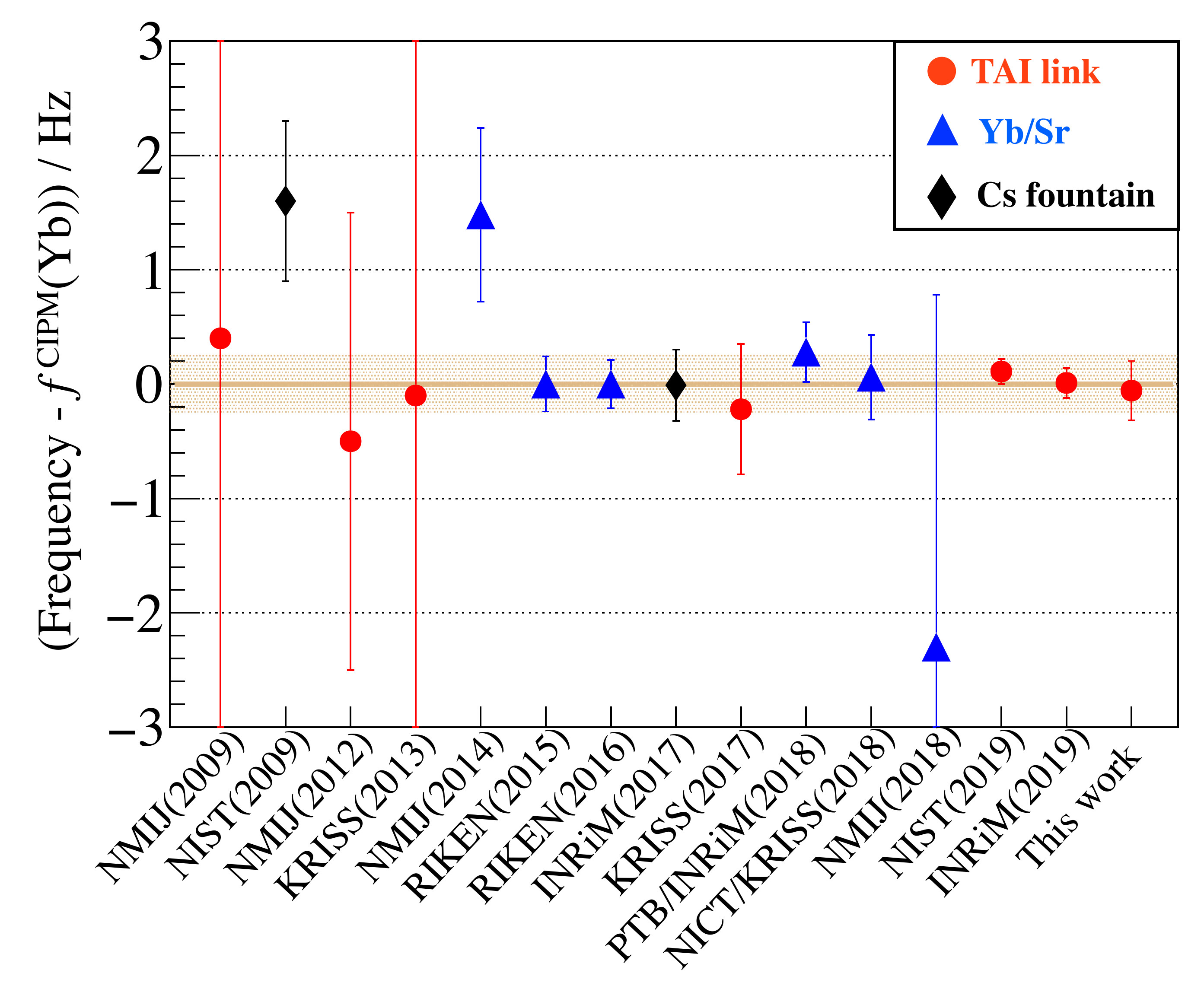}
\caption{Absolute frequencies of the $^{171}$Yb clock transition using a link to TAI \cite{Kohno2009,Yasuda2012,Park2013,Kim2017,McGrew2019,Pizzocaro2019} (red circles), the $^{171}$Yb/$^{87}$Sr frequency ratio \cite{Akamatsu2014OE,Akamatsu2018,Takamoto2015,Nemitz2016,Grotti2018,Fujieda2018} and the CIPM recommended frequency of $^{87}$Sr \cite{Riehle2018} (blue triangles), and local Cs fountain clocks \cite{Lemke2009,Pizzocaro2017} (black diamonds). The shaded region indicates the CIPM recommended value of $^{171}$Yb and $f^{\mathrm{CIPM}}(\mathrm{Yb})=518\,295\,836\,590\,863.6(26)$ Hz \cite{Riehle2018}. NIST: National Institute of Standards and Technology, KRISS: Korea Research Institute of Standards and Science, INRiM: Istituto Nazionale di Ricerca Metrologica, PTB: Physikalisch-Technische Bundesanstalt, NICT: National Institute of Information and Communications Technology.}
\label{comparison}
\end{figure} 

Table \ref{clockbudget} summarizes the uncertainties for the absolute frequency measurement. The statistical uncertainty of the Yb clock was calculated using the stability of the Yb/Sr ratio measurement (see Fig. \ref{allandeviation}) with the duration of the operation. The gravitational red shift was estimated to be $22.9(6)\times10^{-16}$ using the Yb height from the conventionally adopted geoid
potential (62 636 856.0 m$^{2}$/s$^{2}$). The total statistical uncertainty in the full period was $1.8\times10^{-16}$ (dashed blue lines in Fig. \ref{reprocibility}), which was obtained by inflating the uncertainty of the weighted mean by the square root of the reduced chi-square $\sqrt{\chi^{2}_{\mathrm{red}}}=1.3$. Finally, we obtained $y(\mathrm{Yb}-\mathrm{SI})=-1.1(5.0)\times10^{-16}$. The absolute frequency of the Yb clock transition was determined as $518\,295\,836\,590\,863.54(26)$ Hz. The total uncertainty was mostly limited by the systematic uncertainties of the Yb clock and the microwave synthesis. Our determined value agreed with the recommended frequency $f^{\mathrm{CIPM}}(\mathrm{Yb})$ and the results of previous measurements by many groups (see Fig. \ref{comparison}).   

\section{Discussions and conclusion}
The uncertainty in the link between the Yb clock and the SI second was $2.4\times10^{-16}$, which was calculated by using the quadratic sum of the uncertainties due to the dead times in both Yb - UTC(NMIJ) and TAI - SI, the satellite link, and the microwave synthesis. The reduction of the link uncertainty by the continuous operation also enables measurements of the frequency ratios at the $10^{-16}$ level between remotely-located secondary representations of the second. For example, during the present half-year period, a Rb fountain clock (SYRTE FORb) \cite{Guena2014} was running. The frequency difference data $y(\mathrm{TAI}-\mathrm{Rb})$ for 180 days were obtained from Circular T. With our data $y(\mathrm{Yb}-\mathrm{TAI})$, we determined the Yb/Rb ratio as $75\,833.197\,545\,114\,174(42)$ with a relative uncertainty of $5.5\times10^{-16}$. This agreed with the ratio calculated from the recommended frequencies $f^{\mathrm{CIPM}}(\mathrm{Yb})/f^{\mathrm{CPIM}}(\mathrm{Rb})=75\,833.197\,545\,114\,196(59)$ \cite{Riehle2018} and the value $75\,833.197\,545\,114
192(33)$ reported by another group \cite{McGrew2019}. As another example, a Sr optical lattice clock (NICT-Sr1 \cite{Hachisu2018}) calibrated the TAI frequency for 20 days during the sixth campaign period. This enabled us to determine the Yb/Sr ratio as $1.207\,507\,039\,343\,337\,82(75)$ with a fractional uncertainty of $6.2\times10^{-16}$. This was in agreement with the most accurate previous value $1.207\,507\,039\,343\,337\,749(55)$ \cite{Nemitz2016} and the other values with uncertainties at the $10^{-16}$ level \cite{Grotti2018,Takamoto2015,Fujieda2018,Pizzocaro2019}. 

\begin{figure}[h]
\includegraphics[scale=0.4]{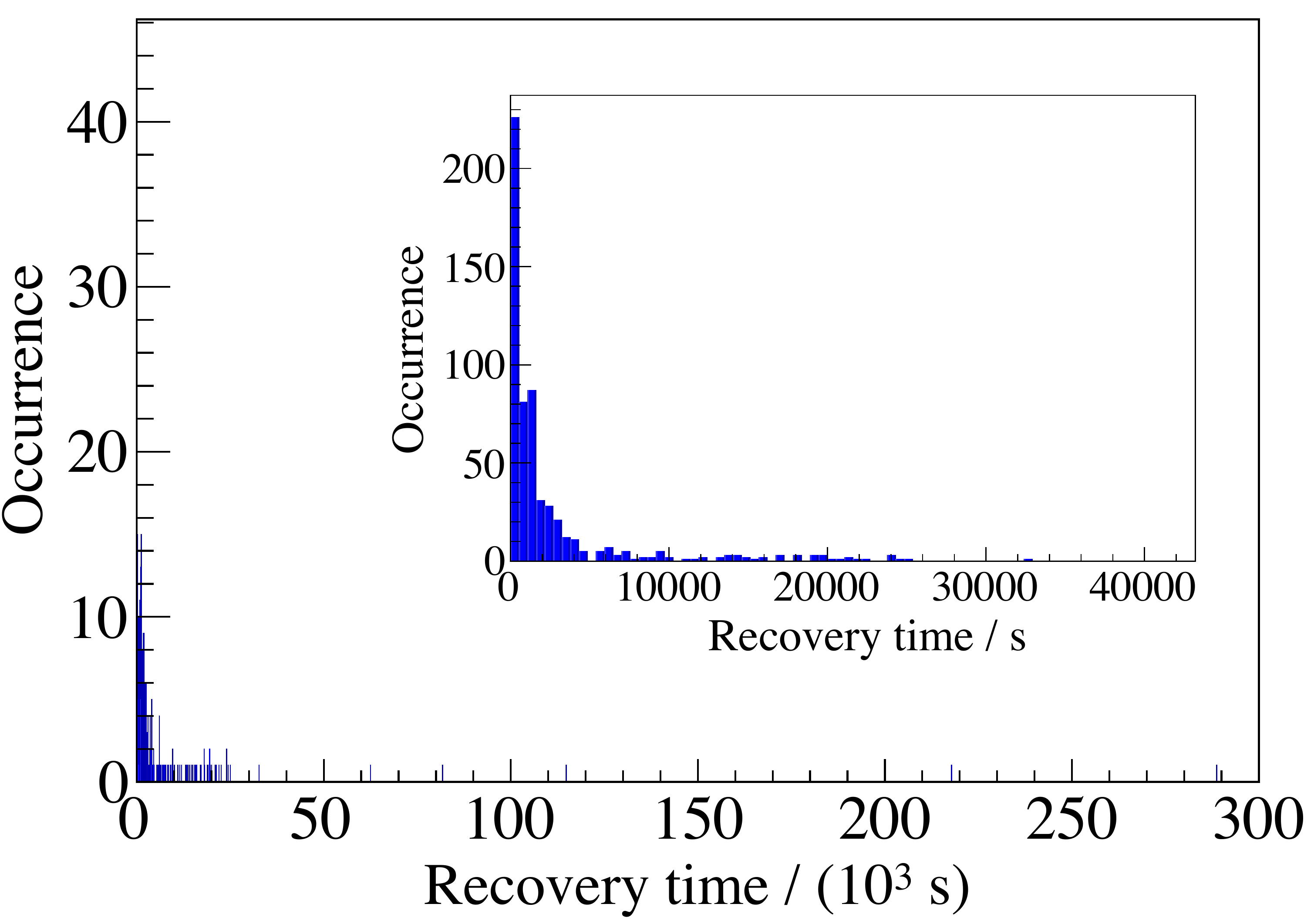}
\caption{Histogram of the recovery time of more than 1 minute. The inset shows recovery times within half a day with a different bin size, from which the mean recovery time was obtained as 0.72 hours.}
\label{histgram}
\end{figure}

The expectation value of the uptime of the Yb clock was calculated under the condition where the Yb clock is fully unattended during a night interval of 8 hours, and where long downtimes resulting from rare events such as facility maintenance are not included. The number of interrupting events that required manual recovery was $580$ during the full campaign period. This was obtained by counting the number of time gaps of more than 1 minute in the frequency data (Fig. \ref{continuousresult}). Since the Yb clock was operated with a $80.3\%$ uptime for 185 days, the mean number of interruptions per hour was $580/(0.803\times185\times24\,\mathrm{hours})=0.16$/hour. To estimate the probability of the occurrence of $n$ interrupting events in an interval of $T$ hours, we employed a Possison distribution,
\begin{equation}
P(n,T)=\frac{(0.16T)^{n}e^{-0.16T}}{n!}.
\label{possion}
\end{equation}
To calculate the expected uptime, the typical time for manually restarting the operation is needed. Figure \ref{histgram} shows a histogram of the recovery time, which was obtained from time gaps in the frequency data (Fig. \ref{continuousresult}). Excluding long gaps exceeding half a day, the mean recovery time was $0.72$ hours. The expected dead time during a 16-hour period of human work was then calculated as $\sum_{n=0}^{\infty}nP(n,16)\times0.72\,{\mathrm{hours}}=1.8$ hours. During an 8-hour night period, the operation is stopped after one interrupting event. In this case, the expected dead time is calculated as $\int_{0}^{8}\frac{d(1-P(0,T))}{dT}(8-T)dT=3.5$ hours. The uptime per day was therefore expected to be $78\%$. 

\begin{figure}[h]
\includegraphics[scale=0.4]{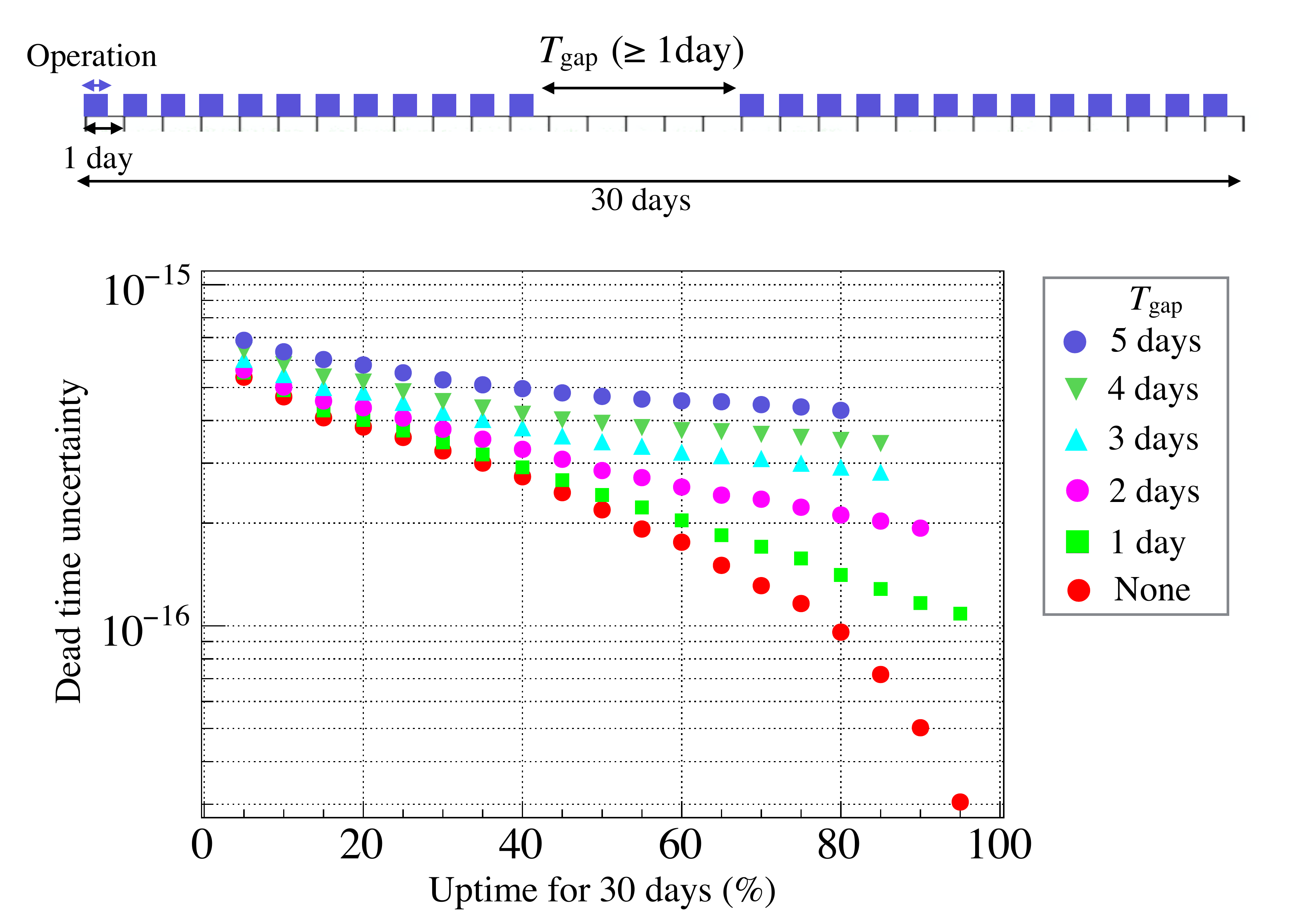}
\caption{Simulated dead time uncertainty as a function of the clock uptime for 30 days. As shown in the top figure, the dead time is homogeneously distributed over the 30-day period, except for a large gap time of $T_{\mathrm{gap}}=1-5$ days in the middle of the period.}
\label{deadtimeplot}
\end{figure}

To consider the optimum conditions for the operation, we carried out a simulation of the dead time uncertainty as a function of the clock uptime (see Figure \ref{deadtimeplot}). In this simulation, the dead time was homogeneously distributed for a 30-day period, except for a large time gap of $T_{\mathrm{gap}}=1-5$ days in the middle of the period. Without the large gap (indicated as “None” in Fig. \ref{deadtimeplot}), the dead time uncertainty falls below $10^{-16}$ for uptimes of $\ge80\%$. When the clock operation is stopped for $T_{\mathrm{gap}}=1-5$ days, the uncertainty no longer reaches below $10^{-16}$ even though the uptime for the overall period is $\ge80\%$. This indicates that the uncertainties calculated with the actual duration of the clock operation are mostly determined by the instabilities of UTC(NMIJ) during the large gaps. When we take into account the fact that the uncertainty of the absolute frequency measurement is limited by the systematic uncertainties of the Yb clock and the microwave synthesis (the combined uncertainty is $4.6\times10^{-16}$), it may be enough to reduce the dead time uncertainty to $\sim2\times10^{-16}$. In this case, the optimum condition is to operate the clock every day with a moderate uptime of $\sim50\%$. This is well below our expectation value of the uptime ($78\%$, see above). It should be noted that daily repetition of the startup and shutdown of the clock increases human effort, and thus it is better in practice to let the clock run continuously. 

Previous studies have demonstrated an improvement in local timescales by steering flywheel oscillators to intermittently operated lattice clocks \cite{Grebing2016,Hachisu2018,Milner2019,Yao2019}. In these methods, the stabilities of the timescales are limited by the fluctuations of the flywheels during the dead time of the clocks, and thus can be further improved by continuous operation. An analysis for our local timescale will be presented in a future publication.

The performance at the present level can make a significant contribution to the search for new physics. For example, local position invariance has been tested by looking for an annual variation of the gravitational red shift of atomic clocks \cite{Blatt2008,Targat2013,McGrew2019}. As demonstrated in previous experiments that utilized continuously running fountain clocks and hydrogen masers \cite{Peil2013}, the continuous operation contributes to reduce the statistical uncertainty in a relatively short term, resulting in a stringent limit on violation of local position invariance. In our case, this experiment can be carried out by comparing the Yb clock with a local Cs fountain clock \cite{Takamizawa2015} or a hydrogen maser. Another example is that transient variations of the fine structure constant resulting from the passage of dark matter may be observed in a global network of optical clocks \cite{Derevianko2014}. The detection of the transient effects relies on correlations in the frequency data of worldwide clocks running simultaneously \cite{Wcislo2018}. To obtain a large amount of temporally overlapped data from laboratories worldwide, it is highly desirable that the participating clocks are capable of running continuously for a long period. The continuous operation also makes it possible to search for dark matter interactions on long timescales that could not be studied in previous experiments \cite{Wcislo2018,Roberts2019}.

In conclusion, we have demonstrated the nearly continuous operation of an Yb optical lattice clock with an uptime of $80.3\%$ for half a year. This enabled us to link the Yb clock to the SI second with an uncertainty of $2.4\times10^{-16}$, and to determine the absolute frequency of the Yb clock transition with a total uncertainty of $5.0\times10^{-16}$. The present performance preliminary demonstrated that the robustness of optical lattice clocks can reach a level comparable to that of Cs fountain clocks. This work constitutes an important step towards the redefinition of the SI second. 

\section*{Acknowledgments}
We thank H. Katori, M. Takamoto, and H. Imai for providing information on their vacuum systems, and S. Okubo for technical assistance. We are indebted to national metrology institutes for their efforts in operating the primary and secondary frequency standards and making the data available in Circular T. This work was supported by Japan Society for the Promotion of Science (JSPS) KAKENHI Grant Number 17H01151, and 17K14367, and JST-Mirai Program Grant Number JPMJMI18A1, Japan.

\end{document}